
\input phyzzx
\def\mo{{\bar 1}}

\def\ot{{1\over 3}}
\def\tt{{2\over 3}}
\def\ft{{4\over 3}}
\def\p32{{\tilde +}}
\def\m32{{\tilde -}}

\hskip 11.4cm SNUTP 92--88
\vskip 1.5cm
\centerline{\fourteenbf An Orbifold Compactification with}
\centerline{\fourteenbf Three Families from Twisted Sectors}
\vskip 2cm
\centerline{Hang Bae Kim and Jihn E. Kim}

\centerline{\sl Center for Theoretical Physics and Department
of Physics}
\centerline{\sl Seoul National University}
\centerline{\sl Seoul 151-742, Korea}
\vskip 2cm
\abstract{We obtain a three generational $SU(3)_c\times SU(3)_w
\times U(1)^4\times   [SO(12)\times U(1)^2]^\prime$
model from an orbifold construction with the requirement that
three  generations arise from twisted sectors.  There exist
supersymmetric vacua realizing the standard model.  In one
example the anomalous $U(1)$ breaks the gauge symmetry down
to $SU(3)_c\times SU(2)_w\times U(1)_Y\times
SO(12)^\prime$.}
\endpage

The orbifold construction of four dimensional string models [1]
has attracted a great deal of attention due to its possibility of
obtaining a standard-like model [2] and the relative simplicity
of model building compared to the other constructions [3,4,5].
In order to achieve a string derived standard model, however,
several phenomenologically desirable features have to be
realized: $(i)$ three families, $(ii)$ a successful hypercharge
assignment, $(iii)$ a good prediction of sin$^2\theta_{\rm W}$
at electroweak scale, $(iv)$ a good prediction of mixing
angles in the quark sector, and $(v)$ a strong CP solution.
In this paper, we
focus on $(i)$ and $(ii)$ with an eye on a possible solution
of $(iv)$.

An important ingredient toward a superstring standard model
is the role of the hidden sector for supersymmetry breaking [6].
A hidden sector confining group is better to be present.  A
desirable hidden sector is two hidden sector
confining groups with
a comparable group size for a realistic determination of gauge
coupling constant at the string scale [7].  In Ref. [7], a hidden
sector with $SU(10)\times SU(9)$ has been considered.  This large
hidden sector cannot be obtained in symmetric orbifold

constructions.  The reason for considering the large
hidden sector gauge group has
been to obtain almost the same but different $\beta$ functions.
This condition for $\beta$ function can be achieved by {\it two
same size groups} but different numbers of hidden matter
fields.  The hidden sector $E_8^\prime$ can have
$SU(5)^\prime \times SU(5)^\prime$ as a candidate
subgroup in this scenario; but this cannot
be obtained by our application of shift vectors.   Therefore,
the largest hidden sector factors with comparable sizes
is  $SU(4)^\prime  \times SU(4)^\prime$.  Our model, however,
does not realize this factor group but one hidden confining
group $SO(12)^\prime$.
Nevertheless, we may anticipate the comparable but slightly
different $\beta$ functions if one $SO(12)^\prime$
is broken to $SU(4)^\prime\times SU(3)^\prime$ at a somewhat
lower scale than the string scale by vacuum expectation values of
Higgs fields.

Another motivation for building a 4-dimensional superstring model
is to understand the flavor problem.  The three generation
superstring standard models proposed so far derive the $SU(2)$
doublets of quarks from the untwisted sector [2]; thus realistic
fermion mass spectrum cannot be obtained. To understand the flavor
problem, the quark doublets must arise from twisted sectors.
This condition is very restrictive in the orbifold construction
of 4-dimensional superstring models. With this condition one
cannot obtain
$SU(3)\times SU(2)\times U(1)$ models or $SU(5)\times U(1)$
models.  The smallest group constructed in this
way is $SU(3)\times SU(3)\times U(1)$'s.

We will consider a model with two Wilson lines to obtain
multiplicity of three automatically.
The shift vector and Wilson lines are
$$\eqalign{v\ &=\ ({\tenrm
\ot\ \ot\ \tt\ \ot\ \ot\ \tt\ 0\ 0)(0\ 0\ 0\
0\ 0\ 0\ 0\ 0})
\cr a_1\ &=\ ({\tenrm
0\ 0\ 0\ \ot\ \ot\ \tt\ \ot\ \ot )(\ot\ \ot\ \ot\
\ot\ 0\ 0\ 0\ 0})
\cr a_3\ &=\ ({\tenrm
0\ 0\ 0\ 0\ 0\ 0\ 0\ \tt )(\ot\ \ot\ \ot\ \ot\ \ot\
\ot\ \ot\ \ot }).}\eqno (1) $$
from which we obtain the desired gauge group
$$SU(3)_c\otimes SU(3)_w\otimes [U(1)]^4\otimes SO(12)^\prime
\otimes [U(1)^\prime ]^2. \eqno (2)$$
Massless chiral fermions arise from untwisted and twisted
sectors.  The fermions in the twisted sectors have opposite
chirality  from  the fermions in the untwisted sector.   The
massless chiral superfields satisfy

$\underline{for\ untwisted\ sector}$
$$ \eqalign{p_I\cdot v\ &=\ -{1\over  3}\
{\rm mod\ 1},\cr
 p_I\cdot a_i\ &=\
0\ \ {\rm mod\ 1},\ \ \ (i=1,3)}\eqno (3)$$

$\underline{for\ twisted \ sectors}$
$$ p_I\cdot p_I=\Bigg\{ \matrix{\tt \ \
({\rm multiplicity\ 9})\cr
\ft\ \  ({\rm multiplicity\ 3}).\cr} \eqno (4)$$
If we add one more Wilson line, the multiplicities in the twisted
sectors become 3 and 1 for $p_I^2=\tt$, and $\ft$, respectively.
In this sense, two Wilson line models are most attractive because
all chiral fields appear as multiples of 3.

There are nine twisted sectors
distinguished by the shift and Wilson lines,
$$\eqalign{
&T0:\ v,\ \ \ T1:\ v+a_1,\ \ \ T2:\ v-a_1,\ \cr
&T3:\ v+a_3,\ \ \ T4:\ v-a_3,\ \ \ T5:\ v+a_1+a_3,\cr
&T6:\ v+a_1-a_3 \ \ \ T7:\ v-a_1+a_3 \ \ \ T8:\ v-a_1-a_3
}\eqno (5)$$

For the twisted sector $T0$, we present all massless chiral
fields.   We drop the multiplicity 3 throughout
the paper.\foot{For two  Wilson lines,
the  original  27  degenerate states of the  twisted  sector  are
distinguished by 9 different but triply degenerate states.}
The momenta satisfying
$p_I^2=\tt$ are
$$\eqalign{p_I\ &=\ (0\ 0\ \mo\ 0\ 0\ \mo\ 0\ 0)(\cdots )\ \
\ \ \ \ 3\cdot {\bf 1}\cr
p_I\ &=\ (-\ -\ -\ -\ -\ -\ +\ +)(\cdots )\ \
\ \ \ \ 3\cdot {\bf 1}\cr
p_I\ &=\ (-\ -\ -\ -\ -\ -\ -\ -)(\cdots )\ \
\ \ \ \ 3\cdot {\bf 1}}\eqno (6)$$
where $\mo, +$, $-$, and $\m32$ represent --1, 1/2, --1/2 and
--3/2, respectively.  The momenta satisfying $p_I^2=\ft$ are
$$\left(\matrix{0\ 0\ 0\ 0\ 0\ 0\ 0\ 0\cr
0\ \mo\ \mo\ 0\ 0\ 0\ 0\ 0\cr
\mo\ 0\ \mo\ 0\ 0\ 0\ 0\ 0\cr
0\ 0\ 0\ 0\ \mo\ \mo\ 0\ 0\cr
0\ \mo\ \mo\ 0\ \mo\ \mo\ 0\ 0\cr
\mo\ 0\ \mo\ 0\ \mo\ \mo\ 0\ 0\cr
0\ 0\ 0\ \mo\ 0\ \mo\ 0\ 0\cr
0\ \mo\ \mo\ \mo\ 0\ \mo\ 0\ 0\cr
\mo\ 0\ \mo\ \mo\ 0\ \mo\ 0\ 0\cr}\right)
=(3_c^*,3_w^*)\eqno (7.a)$$
$$\left(\matrix{\mo\ \mo\ \mo\ 0\ 0\ \mo\ 0\ 0\cr
\mo\ 0\ 0\ 0\ 0\ \mo\ 0\ 0\cr
0\ \mo\ 0\ 0\ 0\ \mo\ 0\ 0\cr}\right)=(3_c,1),\ \
\left(\matrix{0\ 0\ \mo\ \mo\ \mo\ \mo\ 0\ 0\cr
0\ 0\ \mo\ \mo\ 0\ 0\ 0\ 0\cr
0\ 0\ \mo\ 0\ \mo\ 0\ 0\ 0\cr}\right)=(1,3_w)\eqno (7.b)$$
$$\left(\matrix{+\ -\ -\ -\ -\ -\ +\ -\cr
-\ +\ -\ -\ -\ -\ +\ -\cr
-\ -\ \m32\ -\ -\ -\ +\ -\cr}\right)=(3_c,1),\ \
\left(\matrix{+\ -\ -\ -\ -\ -\ -\ +\cr
-\ +\ -\ -\ -\ -\ -\ +\cr
-\ -\ \m32\ -\ -\ -\ -\ +\cr}\right)=(3_c,1)\eqno (7.c)$$
$$\left(\matrix{-\ -\ -\ +\ -\ -\ +\ -\cr
-\ -\ -\ -\ +\ -\ +\ -\cr
-\ -\ -\ -\ -\ \m32\ +\ -\cr }\right)=(1,3_w),\ \
\left(\matrix{-\ -\ -\ +\ -\ -\ -\ +\cr
-\ -\ -\ -\ +\ -\ -\ +\cr
-\ -\ -\ -\ -\ \m32\ -\ +}\right)=(1,3_w)
 \eqno (7.d)$$
where we neglected the eight zero entries of $E_8^\prime$.
We also showed the representation content  in the
gauge group $SU(3)_c\times SU(3)_w$.
Similarly, we find the other chiral fields
which are,
taking the opposite chiralities of the twisted and
untwisted sectors and ignoring the multiplicity 3,
$$\eqalign{&UT:\ (3_c^*,1),\ (1,3_w^*)\cr
&T0:\ (3_c, 3_w),\ 3(3_c^*,1),\ 3(1,3_w^*),\ 9\cdot {\bf 1}\cr
&T1:\ (3_c^*,1),\ (1,3_w),\ 3\cdot {\bf 1}\cr
&T2:\ (3_c,1),\ (1, 3_w^*),\ 3\cdot {\bf 1}\cr
&T3:\ (3_c,1),\ (3_c^*,1),\ (1,3_w),\
(1,3_w^*),\ 6\cdot {\bf 1}\cr
&T4:\ (3_c,1),\ (3_c^*,1),\ (1,3_w),\ (1,3_w^*),\
6\cdot {\bf 1}\cr
&T5:\ (3_c,1),\ (3_c^*,1),\ (1,3_w),\ (1,3_w^*),\
6\cdot {\bf 1}\cr
&T6:\ (12)^\prime ,\ 6\cdot {\bf 1}\cr
&T7:\ (3_c,1),\ (1,3_w),\ 3\cdot {\bf 1}\cr
&T8:\  (3_c,1),\ (3_c^*,1),\ (1,3_w),\ (1,3^*_w),\  6\cdot {\bf
1}.
}\eqno (8) $$
Striking out vectorlike combinations under $SU(3)_c\times
SU(3)_w\times SO(12)^\prime$, we obtain
a three generation model
$$(3_c,3_w)+3(3_c^*,1)+3(1,3_w^*)+{\rm singlets}\eqno (9)$$
where the multiplicity 3 is not written.  Of course, the charged
lepton singlets are hidden in the  singlets.

The six $U(1)$ charges are defined as
$$\eqalign{Q_1&= (1\ 1\ {-1}\ 0\ 0\ 0\ 0\ 0)(0\ \cdots\ 0)\cr
Q_2&=(0\ 0\ 0\ 1\ 1\ {-1}\ 0\ 0)(0\ \cdots\ 0)\cr
Q_3&=(0\ 0\ 0\ 0\ 0\ 0\ {-1}\ 0)(0\ \cdots\ 0)\cr
Q_4&=(0\ 0\ 0\ 0\ 0\ 0\ 0\ {-1})(0\ \cdots\ 0)\cr
Q_5&=(0\ \cdots\ 0)({-1}\ {-1}\ {-1}\ {-1}\ 0\ 0\ 0\ 0)\cr
Q_6&=(0\ \cdots\ 0)(0\ 0\ 0\ 0\ {-1}\ {-1}\ {-1}\ {-1})
} \eqno (10)$$

It is tedious but straightforward to calculate the $Q$ charges
of the matter superfields.  We define new
$U(1)$ charges as $P_1, P_2, P_3, P_4, P_5$, and
$X$,
$$\eqalign{&P_1={1\over 6}(Q_1+Q_2)\cr
&P_2={1\over 2}(Q_3+Q_4)\cr
&P_3={1\over 12}(Q_1-Q_2)+{1\over 4}(Q_3-Q_4)\cr
&P_4={1\over 12}(Q_1-Q_2)-{1\over 12}(Q_3-Q_4)-{1\over 6}Q_6\cr
&P_5={1\over 4}Q_5\cr
&X={1\over 12}(Q_1-Q_2)-{1\over 12}(Q_3-Q_4)+{1\over 12}Q_6
}\eqno (11)$$
The trace of $X$ charge is
$$\eqalign{
\sum_{i=3_c,3_c^*}&X(i)=-2,\ \ \sum_{i=3_w,3_w^*}
X(i)=-2,\ \ \sum_{i=12^\prime}X(i)=-1,\cr\cr
&\sum_{i={\rm all\ including\ singlets}}X(i)=-24
} \eqno (12)$$

This compactification exhibits the anomalous $U(1)_X$
whose gauge boson becomes massive by
absorbing the model independent axion $a_{MI}$.  $X$ is
the charge of this $U(1)_X$.  Note that the divergence of
the corresponding current is
$$\partial^\mu  J^X_\mu=+\{F_c\tilde F_c\}+\{F_w\tilde  F_w\}
+\{F^\prime\tilde F^\prime\}+\cdots
\eqno (13)$$
where $\{F_c\tilde F_c\}\equiv (1/32\pi^2)F^a_{\mu\nu}\tilde
F^{a\mu\nu}$ ($a=SU(3)_c$ adjoint index), etc., and $\cdots$
denote  $U(1)\ F\tilde F$'s with the same coefficient +1.   Even
though the sum of $X$ charges for color triplets and
antitriplets differs from the $X$ charge of $12^\prime$
by a factor of 2, the anomaly coupling given above is the same
because the indices of the representations differ by a factor
of   $\half$,   viz.  $l(3_c\  {\rm  or}  \   3_c^*)=\half$   and
$l(12^\prime)=1$ where Tr $T_iT_j=l\delta_{ij}$.   Thus the
anomaly coefficient of $U(1)_X$ match those of the
model-independent axion.  The model-independent axion becomes
the longitudinal degree of the $U(1)_X$ gauge boson.

The presence of the anomalous $U(1)_X$ has desirable
Fayet--Iliopoulos $D$-terms, reducing the rank of the
gauge group.  For supersymmetry, we must satisfy [8,9]
$$\langle D^{(X)}\rangle
=\langle {2g\over 192\pi^2}{\rm Tr}X
+\sum_iX(i)\phi^*(i)\phi(i)\rangle =0
\eqno (14)$$
One can find vacua satisfying the above supersymmetry condition.
One may reduce the rank of the gauge group by the condition of
vanishing $D$-terms,
and breaking $SU(3)_w$ down to $SU(2)_w$.  One must check also
that the resulting three generation model has the correct
hypercharge.  We find several models satisfying this criteria.

For example, giving a vacuum expectation values to the following
$3_w, 3_w^*$ and singlets,
$$ \eqalign{&\ \ \ \ \ \ P_1\ \ \  P_2\ \ \ \ P_3
\ \ \ P_4\ \ \ \ P_5\ \ \ X\cr
{\bf 3}\ (T0)\ &:\ \ \ \ \  0\ \ \ \
\ 0\ \ \ \ \ 1\ \ \ \ \ 1\ \ \ \ \ 0\ \ \ \ \ 1\cr
{\bf 3^*} (T7)\ &:\ \ \ \ \ 0\ \ \ \ \ 0\ \ \ {-1}
\ \ \ \ \ 1\ \ \ \ \ 0\ \ \  {-1}\cr
{\bf 1}\ (T0)\ &:\ \ \ {-1}\ \ \ \ \ 3\ \ \ \ \ 0\
\ \ \ \ 0\ \ \ \ \ 0\ \ \ \ \ 0\cr
{\bf 1}\ (T0)\ &:\ \ \ {-1}\ \ \ {-3}\ \ \ \ \ 0\ \
\ \ \ 0\ \ \ \ \ 0\ \ \ \ \ 0\cr
{\bf 1}\ (T6)\ &:\ \ \ \ \ 0\ \ \ \ \ 0\ \ \ \ \ 0\ \
\  {-2}\ \ \ \ \ 0\ \ \ \ \ 0\cr
{\bf 1}\ (T6)\ &:\ \ \ \ \ 0\ \ \ \ \ 0\ \ \ \ \ 0\
\ \ \ \ 0\ \ \ \ \ 0\ \ \ \ \ 2\cr
{\bf 1}\ (T6)\ &:\ \ \ \ \ 0\ \ \ \ \ 0\ \ \ \ \ 0
\ \ \ \ \ 0\ \ \ \ \ 3 \ \ \ {-1}\cr
{\bf 1}\ (T6)\ &:\ \ \ \ \ 0\ \ \ \ \ 0\ \ \ \ \ 0
\ \ \ \ \ 0\ \ \  {-3}\ \ \ {-1}\cr
{\bf 1}\ (T7)\ &:\ \ \ \ \ 1\ \ \ \ \ 3\ \ \ \ \ 0\ \
\ \ \ 2\ \ \ \ \ 0\ \ \ \ \ 0\cr
{\bf 1}\ (T7)\ &:\ \ \ \ \ 1\ \ \ {-3}\ \ \ \ \ 0\ \
\ \ \ 2\ \ \ \ \ 0\ \ \ \ \ 0
}\eqno (15)$$
we obtain the desired electroweak hypercharge
$$Y=Y_3+{1\over 3}P_3,\eqno (16)$$
where $Y_3$ is proportional to the 8$^{th}$ generator of
$SU(3)_w, \ \ Y_3=$diag.(1/6,1/6,--1/3).
Thus we obtain the supersymmetric standard model $SU(3)\times
SU(2)\times U(1)_Y\times SO(12)^\prime$.
All the other $U(1)$'s are broken.\foot{If we want an extra
$U(1)$, we can remove (-1-3 0 0 0 0) (T0) and (1 3 0 2 0 0) (T7)
fields in Eq. (16).  Then the additional $U(1)$ gauge charge is
$Y^\prime=3P_1+P_2$ which can
be used to guarantee a long proton lifetime.}
Removing the vectorlike representations, we obtain
three families at low energy.  Among vectorlike representations,
there appear $Q_{em}=\pm 1/3, \pm 2/3$ leptons and $Q_{em}
=\pm 1/3$ quarks.

The compactification realized above hints a few interesting
directions toward string derived supersymmetric standard
models.  Firstly, the standard model gauge group can arise
through the Fayet--Iliopoulos mechanism even though
the 4-D string model possess a much larger gauge symmetry.
This Fayet--Iliopoulos mechanism is like the Higgs mechanism,
and the available Higgs fields are restricted.  A grand
unification such as $SU(5)\times U(1)$ can be broken down
to the standard model gauge group by the Fayet--Iliopoulos
mechanism, but a model with a much
larger gauge symmetry may run into the difficulty due to the lack
of needed Higgs fields.  In the present case, the gauge
group is small enough, $SU(3)\times SU(3)\times U(1)$'s;
there exist a number of needed Higgs fields realizing
supersymmetric standard model.  Second, the requirement
of quark doublets from twisted sectors is very restrictive.  For
example,  it  requires that there should
exists a $\tilde v$\foot{$\tilde  v$  is  the
vector signifying the twisted sector, e.g. $v+a_1$ for $T1$.}
taking the form,\foot{Or even number of 1/3 entries can be
replaced by 2/3's. Also, -- signs and appropriate number of
integers can be added.}

$$\tilde v=(\ot\ \ot\ \ot\ \ot\ \ot\ \ot\ 0\ 0)(\cdots) $$
where $\cdots$ are zeros.  Then, it is easy to see that
$SU(3)\times SU(3)\times U(1)$'s in the observable sector is
the smallest gauge group possible, allowing the
quark doublets.  Third, the anomalous
$U(1)$ allows a low energy global symmetry [10].  However, in the
present example the global symmetry is broken by
the Fayet--Iliopoulos mechanism (viz. the (0 0 0 0 0 2) field in
Eq. (15)), and the axion scale turns
out to be too large.  However, this phenomenon is not universal,
since one may choose a different set of fields for the
Fayet--Iliopoulos symmetry breaking.   Finally, we comment
that models with three Wilson lines do not realize three quark
doublets from the twisted sectors.

In this paper, we find a supersymmetric standard model
$SU(3)_c\times SU(2)_w
\times U(1)_Y\times SO(12)^\prime$.   The three
quark doublets arise from twisted sectors, which is a desirable
feature if the quark mass matrix has to be understood at the
string level.

\vskip 1.5cm
\centerline{\bf Acknowledgments}

This  work is supported in part by Korea Science and  Engineering
Foundation through Center for Theoretical Physics, Seoul
National University.
\endpage
\centerline{\bf References}
\pointbegin L. Dixon, J. Harvey, C. Vafa, and E. Witten, Nucl.
Phys. {\bf B261} (1985) 678; {\bf B274} (1986) 285; L. E. Ibanez,
H. P. Nilles and F. Quevedo, Phys. Lett. {\bf B187} (1987) 25.

\point L. E. Iba$\tilde {\rm n}$ez, J. E. Kim, H. P. Nilles, and
F. Quevedo, Phys.  Lett. {\bf B191} (1987) 282;
L. E. Iba$\tilde {\rm n}$ez, J. Mas,  H.  P. Nilles, and F.
Quevedo, Nucl. Phys. {\bf B301} (1988) 157;
J. A. Casas and C. Munoz, Phys. Lett. {\bf B214} (1988) 63.

\point B. R. Greene, K. H. Kirklin, P. J. Miron, and G. G. Ross,
Phys. Lett. {\bf B180} (1986) 69; Nucl. Phys. {\bf B278} (1986)
667; {\bf B292} (1987) 606.

\point  I.  Antoniadis,  J.  Ellis, J.  S.  Hagelin,  and  D.  V.
Nanopoulos, Phys. Lett. {\bf B208} (1988) 209.

\point D. Gepner, Nucl. Phys. {\bf B296} (1988) 757; Phys. Lett.
{\bf B199} (1987) 380.

\point  J.-P. Derengdinger, L. E. Ibanez and H. P. Nilles,  Phys.
Lett. {\bf B155} (1985) 65; M. Dine, R. Rohm, N. Seiberg, and
E. Witten, Phys. Lett. {\bf B156} (1985) 55.

\point V Kaplunovsky, L. Dixon, J. Louise, and M. Peskin,
SLAC--Pub--5256 (1990).

\point M. Dine, N, Seiberg and E. Witten, Nucl. Phys. {\bf B289}
(1987) 317; J. J. Atick, L. J. Dixon and A. Sen, Nucl. Phys. {\bf
B292} (1987) 109; M. Dine, I. Ichinose and N. Seiberg, Nucl.
Phys. {\bf B293} (1987) 253.

\point A. Font, L. E. Ibanez, H. P. Nilles, and F. Quevedo,
Nucl. Phys. {\bf B307} (1988) 109; J. A. Casas, E. K. Katehou
and C. Mu$\tilde {\rm n}$oz, Nucl. Phys. {\bf B317} (1989) 171.

\point J. E. Kim, Phys. Lett. {\bf B207} (1988) 434.
\bye